\documentclass[12pt,dvips]{article}

\usepackage{amsmath,amssymb,exscale}
\usepackage{array,multicol}
\usepackage{afterpage,float,flafter}
\usepackage{epsfig,rotating,pifont}
\usepackage{cite}
\@addtoreset{equation}{section} \makeatother

\title{
\vspace{1cm}
\Huge\textbf{Footballs, Conical Singularities\\ and the Liouville
Equation}
\vspace*{.5cm}
\author{\large \textbf{Michele Redi\footnote{email:
redi@physics.nyu.edu}}\\
\emph{Department of Physics and CCPP, New York University}\\
\emph{4 Washington Place, New York, NY 10003}}}

\date{}
\begin{document}
\maketitle \thispagestyle{empty} \vspace*{.5cm}

\begin{abstract}
We generalize the football shaped extra dimensions scenario to an
arbitrary number of branes. The problem is related to the solution
of the Liouville equation with singularities and explicit
solutions are presented for the case of three branes. The tensions
of the branes do not need to be tuned with each other but only
satisfy mild global constraints.
\end{abstract}

\newpage
\renewcommand{\thepage}{\arabic{page}}
\setcounter{page}{1}

\section{Introduction}

It is a notoriously difficult problem to find static solutions of
Einstein's equations coupled to brane sources. Exact solutions can
sometimes be found in supergravity theories in the BPS limit but
little is known for non-supersymmetric compactifications.
Codimension two branes are in this regard special. In the simplest
cases, as particles in 2+1 dimensions \cite{djh}, the branes do
not curve the space outside of the source but only create a
deficit angle. The simplified dynamics of gravity then allows to
determine many interesting solutions
\cite{carroll,navarro,codimension2}. Recently codimension two
brane-worlds have also drawn a lot of attention especially in
relation to the cosmological constant problem.

In this note we study generalizations of the so called football
shaped extra dimensions scenario \cite{carroll,navarro} to include
several codimension two branes. Our results can also be repeated
almost verbatim for the Supersymmetric Large Extra Dimensions
scenario \cite{burgess}, which can be considered as a
supersymmetric extension of this model, and more in general for
product compactifications where the internal space is a sphere
(warped compactifications in $6D$ supergravity have also been
considered in \cite{warped}). In \cite{carroll,navarro}, the
authors considered a compactification of six dimensional gravity
to Minkowski space times a sphere, obtained by tuning the magnetic
flux of a $U(1)$ gauge field through the sphere with the bulk
cosmological constant. It was found that by placing \emph{equal}
tension branes at the antipodal points of the sphere the internal
space is deformed into a sphere with a wedge removed (a
"football"). A very interesting feature of this scenario is that
the large dimensions remain flat even in the presence of the
branes. While the tuning between the tensions can be justified
assuming a $\mathbb{Z}_2$ symmetry, certainly this solution
appears very special. It is the purpose of this paper to show that
these types of solutions are quite generic and \emph{no tuning}
between the tensions needs to be invoked when several branes are
considered. The mathematical problem consists in solving the
Liouville equation with singularities, a topic which appears in
$2D$ quantum gravity. Quite remarkably we will be able to find
explicit solutions for the case with three branes but solutions
exist in general. The space so constructed describes a sphere with
conical singularities at the brane locations.

This paper is organized as follows. In section 2 we review our
model and generalize it to an arbitrary number of branes and
curved background. In section \ref{liouvillesec} the problem of
determining the metric on the internal space is related to the
Liouville equation with singularities. Some background material
regarding the solution of the Liouville equation is reviewed in
the appendix. In \ref{3branes} we derive exact solutions for the
metric with three branes. In \ref{morebranes} and \ref{riemann} we
discuss the case where four or more branes are included and
consider the scenario where the internal manifold is a Riemann
surface. We derive the low energy effective action of the model in
section \ref{effectiveaction}. In section \ref{conclusions} we
summarize the results.

\section{The model}
\label{model}

In this section we review and generalize the scenario introduced
in \cite{carroll,navarro}. For appropriate values of the
parameters this is just a truncation of the SLED scenario. The
bulk action is $6D$ gravity with cosmological constant coupled to
a $U(1)$ gauge field,
\begin{equation}
S_6=M_6^4 \int d^6x \sqrt{-G}\left(\frac 1 2 R -\frac 1 4
F^2-\lambda\right)
\end{equation}
The branes are assumed to be minimally coupled and infinitesimal
so their action is just the Nambu-Goto action,
\begin{equation}
S_{branes}=-\sum_{i=1}^N T_i \int d^4 x \sqrt{-g_i}
\end{equation}
where $g_i$ is the induced metric on each brane. Thick branes have
been considered in \cite{thick}.

We will be interested in product compactifications of the form
$M_4 \times K$ where $M_4$ is maximally symmetric and $K$ is a
compact two dimensional manifold. The metric is given by,
\begin{equation}
ds^2=g_{\mu\nu}dx^\mu dx^\nu +\psi(z,\bar{z}) dz d\bar{z}
\label{ansatz}
\end{equation}
where for convenience we have introduced complex coordinates on
the internal manifold. The branes are located at points $z_i$ in
the internal space. Consistently with the equations of motion it
is assumed that the gauge field has a magnetic flux threading the
internal space,
\begin{equation}
F=i \ B_0 \ \psi(z,\bar{z}) \ dz\wedge d{\bar{z}}, \label{flux}
\end{equation}
where $B_0$ is a constant. Using the ansatz (\ref{ansatz}) and
(\ref{flux}) one finds (see \cite{carroll}),\footnote{We use
normalizations where $\int d^2z \delta(z,\bar{z})=1$.}
\begin{eqnarray}
R^4_{\mu\nu}&=&\frac 1 2 \left(\lambda-\frac 1 2
B_0^2\right)g_{\mu\nu}\label{einstein1} \\
\partial_z\partial_{\bar{z}} \log \psi&=&-\frac k  2
\psi-\sum_{i=1}^N \frac {T_i} {M_6^4}
\delta(z-z_i,\bar{z}-\bar{z}_i), \label{einstein2}
\end{eqnarray}
where $k$ is the curvature of the internal manifold,
\begin{equation}
k=\frac \lambda 2 + \frac 3 4 B_0^2.
\end{equation}
Looking at (\ref{einstein1}) we note that a very remarkable thing
has happened: the four dimensional metric does not depend on the
brane sources. The only effect of the branes in the vacuum is to
change the geometry of the internal space without affecting the
vacuum energy of the four dimensional ground state. As pointed out
in \cite{porrati,cline}, however, this should not lead to easy
enthusiasms regarding solutions of the cosmological constant
problem. Eq. (\ref{einstein2}) is the famous Liouville equation
describing a two dimensional metric of constant curvature $k$. We
will study at length this equation and its solutions in the next
section.

Depending on the value of $B_0$ and $\lambda$ the four dimensional
ground state will be de Sitter, anti-de Sitter or Minkowski
space,\footnote{This has also been discussed long ago in
\cite{randjbar}.}
\begin{equation}
\begin{cases}
\lambda> \frac {B_0^2} 2 ~~~~~~~~~dS_4 \\
\lambda< \frac {B_0^2} 2 ~~~~~~~~~AdS_4\\
\lambda= \frac {B_0^2} 2 ~~~~~~~~~M_4
\end{cases}
\label{cases}
\end{equation}
For the Minkowski and de Sitter case one finds that the curvature
$k$ of the internal space is positive. In section \ref{riemann} we
will also consider the case with negative $k$ where the ground
state is AdS. This leads naturally to compactifications on Riemann
surfaces.

In \cite{carroll,navarro} the authors considered the case of a
brane located at $z=0$. Assuming axial symmetry one readily finds
the solution,
\begin{equation}
\psi=\frac {(1-\alpha_1)^2} k \frac
{4(z\bar{z})^{-\alpha_1}}{\big[1+(z\bar{z})^{1-\alpha_1}\big]^2}
\label{2branes}
\end{equation}
where we have defined,
\begin{equation}
\alpha_1=\frac {T_1} {2\pi M_6^4}.
\label{alpha}
\end{equation}
With a simple change of variables one can see that this is just
the metric of a sphere with radius $1/\sqrt{k}$ with a wedge
removed, the football. The deficit angle is $2\pi \alpha_1$ so
clearly $\alpha_1<1$. Physically we will only allow positive
tension branes so we also assume $0<\alpha_1<1$.

The solution (\ref{2branes}) implies the existence of a second
brane with exactly the same tension at $z=\infty$ (the north pole
of the sphere). In fact, up to reparametrization, this is the only
solution (with no warping) with two branes (see appendix). As we
shall show the tuning between the tensions can be removed
considering three or more branes.

\section{Liouville equation}
\label{liouvillesec}

The mathematical problem of determining the metric on the internal
space consists in finding solutions of the Liouville equation with
prescribed singularity on the complex plane,
\begin{equation}
\partial_z \partial_{\bar{z}}\log \psi=-\frac k 2 \psi-2\pi
\sum_{i=1}^N \alpha_i~\delta(z-z_i,\bar{z}-\bar{z}_i)
\label{liouville}
\end{equation}
where the $\alpha_i$'s are related to the tensions as in
(\ref{alpha}). The left hand side of this equation is proportional
to the two dimensional curvature $\sqrt{\gamma} R_2$ of the
internal space. Integrating the Liouville equation and using the
Gauss-Bonnet formula for compact surfaces with no boundaries,
\begin{equation}
\frac 1 {4 \pi}\int \sqrt{\gamma} R_2=2 -2 g
\label{gaussbonnet}
\end{equation}
(where $g$ is the genus of the surface), one derives a simple
formula for the volume,
\begin{equation}
V_2=\frac {2\pi} {k}(2-2g-\sum_i \alpha_i). \label{volume}
\end{equation}
Clearly a compact solution can only exist when $V_2>0$.

For the case of negative curvature $k$ this equation has been
extensively studied starting with the work of Poincar\'e and
Picard, in particular in relation to the problem of uniformization
of Riemann surfaces. The general result is that a unique solution
describing a compact Riemann surface of genus $g$ exists unless it
is forbidden by the volume formula (\ref{volume}) \cite{troyanov}.
Until section \ref{riemann} we will be interested in the positive
curvature case which is relevant for the Minkowski background. To
the best of our knowledge much less is known in this case. In
fact, we will find that an additional constraint on the tensions
applies.

Since we only allow positive tension branes, the positivity of the
volume forces $g=0$ and\footnote{In the special case $k=0$ it is
possible to compactify the space on the topology of the sphere but
the tensions need to be tuned so that $\sum_i \alpha_i=2$
\cite{sundrum}. The metric in this case is easily found to be
given by $\psi=A\, \Pi_i |z-z_i|^{-2 \alpha_i}$ and the volume
remains arbitrary.}
\begin{equation}
\sum_{i=1}^N \alpha_i < 2.
\end{equation}

Away from the singularities the most general solution of the
Liouville equation with positive curvature is given by,
\begin{equation}
\psi=\frac 1 k \frac {4 |w'|^2}{\left[1+|w|^2\right]^2}
\label{solution}
\end{equation}
where $w(z)$ is an arbitrary holomorphic function. For the
simplest case $w=z$ one recognizes (\ref{solution}) as the metric
of the stereographically projected sphere.\footnote{A simple
physical argument suggests the form of the solution
(\ref{solution}). Since codimension two objects locally do not
curve the space, away from the branes the metric must still be the
metric of a sphere. In fact, starting with the metric of the
Riemann sphere and performing the change of variables $z\to w(z)$
one obtains (\ref{solution}).} In terms of the K\"ahler potential
the metric can be derived from,
\begin{equation}
K=\frac 4 k \log[1+w\bar{w}].
\end{equation}

Given that in two dimensions,
\begin{equation}
\partial_z \partial_{\bar{z}}\log |z|^2=2\pi\,\delta(z,\bar{z}) \label{delta}
\end{equation}
the Liouville equation (\ref{liouville}) implies the following
asymptotic behaviors near the singular points,
\begin{eqnarray}
\psi &\sim& |z-z_i|^{-2\alpha_i} ~~~~~\text{as $z\to z_i$}
\nonumber \\
\psi &\sim& |z|^{-2(2-\alpha_{\infty})} ~~~~~~\text{as $z\to
\infty$}.
\end{eqnarray}
Integrability of the metric around the singularities then requires
\begin{equation}
\alpha_i < 1.
\end{equation}
This is equivalent to the statement that the deficit angle around
each singularity cannot exceed $2\pi$. For $\alpha_i \ge 1$
solutions can still be found but they do not describe compact
spaces.

Coming to the main point, the function $w(z)$ reproducing the
prescribed singularities can be found using the technology of the
fuchsian equations which we review in the appendix. In brief,
given $N$ singularities ($z_i$, $\alpha_i$) one considers the
fuchsian equation,
\begin{equation}
\frac {d^2 u} {dz^2}+\sum_{i=1}^{N}\left[\frac
{\alpha_i(2-\alpha_i)}{4(z-z_i)^2}+\frac
{c_i}{2(z-z_i)}\right]u=0. \label{fuch}
\end{equation}
where $c_i$ are known as the accessory parameters. The required
function $w$ is then given by,
\begin{equation}
w(z)=\frac {u_1(z)} {u_2(z)}
\end{equation}
where $u_1$ and $u_2$ are two linearly independent solutions of
(\ref{fuch}) such that their monodromy around the singular points
is contained in $SU(2)$, i.e. $u_1$ and $u_2$ are multivalued
functions on the complex plane and transform with an $SU(2)$
rotation going around the singularities. To see how this formalism
works in practise we now turn to the case with three
singularities. In the appendix the solution with two singularities
is also derived using the technique of the fuchsian equations.

\subsection{Solution with 3 branes}
\label{3branes}

With three branes an explicit solution of the Liouville equation
can be found in terms of hypergeometric functions. Using
reparametrization invariance it is convenient and conventional to
choose the singularities at $(0,1,\infty)$.\footnote{Notice that
the physical position of the singularities does not depend on this
choice.} The relevant fuchsian equation is given by,
\begin{equation}
\frac {d^2 u} {dz^2}+\frac 1 4\left[\frac
{\alpha_1(2-\alpha_1)}{z^2}+\frac
{\alpha_2(2-\alpha_2)}{(z-1)^2}+\frac
{\alpha_1(2-\alpha_1)+\alpha_2(2-\alpha_2)-\alpha_{\infty}(2-\alpha_\infty)}{z(1-z)}\right]u=0.
\end{equation}
To determine solutions with $SU(2)$ monodromies we follow
\cite{ciafaloni} where the same problem for the case of $SU(1,1)$
monodromies was considered (see also \cite{hasadz} for similar
work). Two linearly independent solutions of the previous equation
are,
\begin{eqnarray}
u_1&=&\displaystyle{K_1 \, z^{(1-\frac {\alpha_1}
2)}\,(1-z)^{\frac {\alpha_2} 2}\,
\tilde{F}[a_1,b_1,c_1,z]}\nonumber\\
u_2&=&\displaystyle{K_2 \, z^{\frac{\alpha_1} 2}\,(1-z)^{\frac
{\alpha_2} 2}\,\tilde{F}[a_2,b_2,c_2,z]} \label{solutions}
\end{eqnarray}
where as in \cite{ciafaloni} we found it convenient to define
modified hypergeometric functions,
\begin{equation}
\tilde{F}[a,b,c,z]=\frac {\Gamma[a]\Gamma[b]}{\Gamma[c]} ~
_2F_1[a,b,c,z],
\end{equation}
and the indexes are,
\begin{eqnarray}
a_1&=&\frac {(2-\alpha_1+\alpha_2-\alpha_{\infty})} 2
~~~~~~~~~~~~~~~~~~~~~a_2=\frac {\alpha_1+\alpha_2-\alpha_{\infty}}
2
\nonumber\\
b_1&=&- \frac {(\alpha_1-\alpha_2-\alpha_{\infty})} 2
~~~~~~~~~~~~~~~~~~~~~~~~b_2=\frac
{-2+\alpha_1+\alpha_2+\alpha_{\infty}} 2
\nonumber\\
c_1&=& 2 -\alpha_1 ~~~~~~~~~~~~~~~~~~~~~~~~~~~~~~~~~~~~~~~c_2=
\alpha_1
\end{eqnarray}
Since the hypergeometric functions are regular at the origin (they
have a branch cut between 1 and $\infty$), the monodromy around
$z=0$ is diagonal,
\begin{equation}
\displaystyle{M_0=\hat{M}(\alpha_1) \ = \ \left(\begin{array}{cc}
e^{-i \pi
\alpha_1} & 0 \\
0 & e^{i \pi \alpha_1}
\end{array} \right)}.
\end{equation}
Expanding (\ref{solutions}) around $z=1$ one finds,
\begin{equation}
u_i\sim  a_{i1}(z-1)^{1-\frac {\alpha_2} 2}+ a_{i2} (z-1)^{\frac
{\alpha_2} 2}
\end{equation}
where,
\begin{equation}
\displaystyle{a_{ij}\ =(A)_{ij}\ = } \left(  \begin{array}{cc}
\displaystyle{K_1\, \Gamma(\alpha_2-1)} & \displaystyle{K_1\,
\frac
{\Gamma(1-\alpha_2)\Gamma(a_1)\Gamma(b_1)}{\Gamma(c_1-a_1)\Gamma(c_1-b_1)}}\\
\\
\displaystyle{K_2\, \Gamma(\alpha_2-1)} & \displaystyle{K_2\,\frac
{\Gamma(1-\alpha_2)\Gamma(a_2)\Gamma(b_2)}{\Gamma(c_2-a_2)\Gamma(c_2-b_2)}}
\end{array} \right)
\end{equation}
This allows to compute the monodromy around $z=1$,
\begin{equation}
M_1=A \hat{M}(\alpha_2) A^{(-1)}= \left(
\begin{array}{cc} \cos\pi\alpha_2 - i \displaystyle{\frac
{a_{11}\,a_{22}+ a_{12}\,a_{21}} {a_{11}\,a_{22}-a_{12}\,a_{21}}}
\sin \pi \alpha_2 &
\displaystyle{2 i \frac { a_{11}\,a_{12}}{a_{11}\,a_{22}-a_{12}\,a_{21}}} \sin \pi \alpha_2 \\
\\ \displaystyle{-2 i\frac { a_{21}\,a_{22}}{a_{11}\,a_{22}-a_{12}\,a_{21}}} \sin \pi \alpha_2
 & \cos\pi\alpha_2 + i \displaystyle{\frac
{a_{11}\,a_{22}+ a_{12}\,a_{21}} {a_{11}\,a_{22} -
a_{12}\,a_{21}}} \sin \pi \alpha_2
\end{array} \right)
\end{equation}
In general this is an $SL(2,\mathbb{C})$ matrix. The condition
that the monodromy is contained in $SU(2)$ then boils down to,
\begin{equation}
(M_1)^+_{12}=-(M_1)_{21},
\end{equation}
which determines the ratio $|K_1/K_2|$. A short computation shows,
\begin{equation}
\left|\frac {K_1} {K_2} \right|^2=-\frac
{\Gamma[a_2]\Gamma[b_2]\Gamma[c_1-a_1]\Gamma[c_1-b_1]}
{\Gamma[a_1]\Gamma[b_1]\Gamma[c_2-a_2]\Gamma[c_2-b_2]}= -
\frac{\cos\pi(\alpha_1-\alpha_2)-\cos \pi \alpha_{\infty}}{\cos
\pi(\alpha_1+\alpha_2)-\cos \pi \alpha_{\infty}}
\end{equation}
The expression above is not positive definite for each set
($\alpha_1$, $\alpha_2$, $\alpha_\infty$) that satisfies $\sum_i
\alpha_i<2$. Assuming without loss of generality that
$\alpha_{\infty}\ge \alpha_{1,2}$, the requirement that the right
hand side be positive implies the non trivial constraint,
\begin{equation}
\alpha_\infty < \alpha_1+\alpha_2. \label{extracondition}
\end{equation}
This is an important result as it is independent from the
Gauss-Bonnet formula.\footnote{This restriction agrees with the
result recently found in \cite{eremenko}.} One can also check
using the formulas in \cite{ciafaloni} that the monodromy at
infinity does not give extra constraints. In general this is a
consequence of the fact that,
\begin{equation}
\Pi_i M_i= 1
\end{equation}
Having determined the functions ($u_1$, $u_2$) with $SU(2)$
monodromies, the Liouville equation is solved by $w=u_1/u_2$.

In summary we have shown that a solution for the metric of the
internal space with three branes exists as long as $\sum_i
\alpha_i<2$ and $\alpha_\infty<\alpha_1+\alpha_2$. The solution is
given in terms of the holomorphic function $w$,
\begin{equation}
w(z)=\frac {K_1} {K_2}\, \frac
{\tilde{F}[a_1,b_1,c_1,z]}{\tilde{F}[a_2,b_2,c_2,z]}\,z^{1-\alpha_1}
\label{finalsolution}
\end{equation}
which determines the metric on the Riemann sphere through
(\ref{solution}). Physically when $\alpha_\infty \to
\alpha_1+\alpha_2$ the proper distance between the point $z=0$ and
$z=1$ goes to zero. In this limit the solution then reduces to the
one with two singularities. In fact the condition
(\ref{extracondition}) implies that when only two singularities
are present $\alpha_1=\alpha_\infty$.

\subsection{More branes}
\label{morebranes}

When four or more singularities are included the situation becomes
immediately much more involved. In principle for $N$ singularities
the canonical way to proceed would be to consider the fuchsian
equation (\ref{fuch}). With an $SL(2,C)$ transformation we can
again fix the positions of three singularities at (0, 1, $\infty$)
leaving $N-3$ undetermined. The accessory parameters $c_i$ satisfy
three linear equations (see appendix) so one can express $c_1$,
$c_2$ and $c_\infty$ as linear combinations of
$c_3$,...,$c_{N-1}$. The remaining accessory parameters should
then be determined from the requirement that the monodromy of two
linearly independent solutions of the fuchsian equation
(\ref{fuch}) belongs to $SU(2)$. Counting the number of equations
one sees that the position of $N-3$ singularities remains
unconstrained. In physical terms this means that the physical
position of $N>3$ branes is not fixed; $N-3$ complex moduli label
different vacua. Unfortunately the solution of the fuchsian
equation with more than two singularities (plus the one at
infinity) is not known in closed form so we could not find
explicit solutions. Some progress in this direction was done in
\cite{menotti} where the problem with three finite singularities
and one infinitesimal was solved in the context of $SU(1,1)$
monodromies. The same methods could be applied here.

Besides the problem of finding exact solutions, it would be
important, both from the physical and mathematical point of view,
to determine for which values of $\alpha_i$ a solution of the
Liouville equation with positive curvature exists and is unique.
To the best of our knowledge, contrary to the negative curvature
case, this is not known \cite{eremenko}. With no pretence of
giving a proof here we notice that from the discussion at the end
of the previous paragraph it would seem natural that,
\begin{equation}
\alpha_\infty<\sum_{i=1}^{N-1} \alpha_n, \label{nbranes}
\end{equation}
where we have assumed $\alpha_\infty\ge \alpha_i$. This
generalizes the formula with two and three singularities and
reduces to it when $N-3$ tensions are taken to zero.

\subsection{Riemann Surfaces}
\label{riemann}

We shall now consider compactifications where the internal
manifold has negative curvature (similar compactifications of
string theory have appeared very recently in \cite{silverstein}).
In the model under investigation this corresponds to,
\begin{equation}
\lambda<-\frac 3 2 B_0^2,
\end{equation}
which implies that the four dimensional ground state is AdS$_4$.
In general, starting from a theory in AdS$_{d+3}$ we could
consider compactifications to AdS$_{d+1}\times K$ which might have
interest from the point of view of the AdS/CFT correspondence
\cite{maldacena}.

In absence of singularities, the metric of the internal space is,
\begin{equation}
\psi=-\frac 1 k \frac 4 {\big[1-z\bar{z}\big]^2},~~~~~~~~|z|\,<\,1
\end{equation}
i.e. the hyperbolic metric on the unit disk $D$. This manifold is
non-compact but we can obtain a compact space considering the
coset $D/\Gamma$ where $\Gamma$ is an appropriately chosen
discrete subgroup of the isometries $SU(1,1)$ that acts without
fixed points in $D$. The space so constructed is a compact Riemann
surface of constant negative curvature $k$ and genus $g$.

Including branes leads again to the Liouville equation
(\ref{liouville}) but $k$ is now negative. This is the case most
commonly studied in the literature and a wealth of results is
available (see \cite{takhtajan} and Refs. therein). The general
solution of the Liouville equation with negative curvature is
given by,
\begin{equation}
\psi=-\frac 1 {k} \frac {4|w'|^2}{\big[1-|w|^2\big]}.
\label{solution2}
\end{equation}
The holomorphic function $w(z)$ can in principle be found using
techniques similar to the ones described in section
\ref{liouville}. According to Picard's theorem (and its
generalizations \cite{troyanov}), a solution of the Liouville
equation with negative curvature exists and is unique provided
that the topological constraint (\ref{volume}),
\begin{equation}
\sum_{i=1}^N \alpha_i>(2-2g) \label{topological}
\end{equation}
is satisfied. Curiously deficit angles increase the volume when
the curvature is negative. Notice that the additional condition
$\alpha_\infty<\sum_{i=1}^{N-1}\alpha_i$ that appears when $k$ is
positive is automatically satisfied. It should be mentioned that
in the negative curvature case the singularities $\alpha_i=1$ are
also allowed. These are called parabolic points and play a special
r\"ole due to their relation to the uniformization of Riemann
surfaces. The asymptotic behavior of the metric is,
\begin{equation}
\psi\sim \frac 1 {|z-z_i|^2 (\log|z-z_i|)^2}~~~~~~~~~~\text{as
$z\to z_i$}
\end{equation}
The singularity is integrable so that the volume remains finite.
The proper distance from the singularity to any point at finite
$z$ is however infinite so the space constructed with these
singularities is non compact.

As an example we can consider the case $g=0$, the so called
hyperbolic sphere. This requires at least three singularities such
that $\sum_{i=1}^3 \alpha_i>2$. The fuchsian equation is exactly
the same as the one studied in section \ref{3branes} but we need
to impose that the monodromies belong to $SU(1,1)$. This requires,
\begin{equation}
\left|\frac {K_1} {K_2}\right|^2= \frac{\cos
\pi(\alpha_1-\alpha_2)-\cos \pi \alpha_{\infty}}{\cos
\pi(\alpha_1+\alpha_2)-\cos \pi \alpha_{\infty}}
\end{equation}
By inspection it is not hard to show that the right hand side of
this equation is always positive definite for the allowed values
of $\alpha_i$ so that a solution always exists. The function $w$
is again given by (\ref{finalsolution}).

\section{Effective action}
\label{effectiveaction}

In this section we discuss the low energy effective action valid
at energies smaller than the curvature $k$.

We start by noting that in absence of branes and for positive
curvature the internal space is a sphere whose isometry group is
$SO(3)$. Upon Kaluza-Klein (KK) reduction one obtains an unbroken
$SO(3)$ gauge theory\footnote{As is well known Riemann surfaces do
not possess any continuous isometry so there are no massless KK
gauge bosons from the metric when the curvature is negative.} (for
the detailed KK reduction see \cite{randjbar}). In addition to
this, from the reduction of the $6D$ gauge field one also obtains
an extra $U(1)$ gauge field which however will not play a role in
what follows. Placing equal tension branes at the poles has the
effect of removing a wedge from the sphere. This breaks $SO(3)\to
U(1)$ so that a massless $U(1)$ gauge boson survives. The other
two gauge bosons are Higgsed by the presence of the branes. From
the low energy point of view we can understand this as follows.
Each brane carries two physical degrees of freedom describing the
fluctuations of the brane in the internal space. Two of these
degrees of freedom are precisely the Goldstone bosons necessary to
implement the breaking $SO(3)\to U(1)$ spontaneously. These modes
correspond to the overall rotation of the system. In this language
choosing the singularities at fixed positions (0, $\infty$)
corresponds to the unitary gauge. The remaining two degrees of
freedom describe the relative motion of the branes. These modes
are massive as the branes repel from each other. When the third
brane is added the original $SO(3)$ symmetry is completely broken.
Out of the two new degrees, the one describing the rotation around
the axis is eaten by the $U(1)$ gauge boson while the other is
massive (this is implied by the fact that the distance between the
branes is fixed in the vacuum). Adding more branes obviously does
not change this picture for the gauge bosons but introduces new
massless degrees of freedom. As we have seen in section
\ref{morebranes}, for $N>3$ the physical positions of the branes
is not determined in the vacuum and they will appear as $N-3$
complex flat directions of the potential in the low energy
effective theory. An interesting object to consider in this case
would be the metric on the moduli space. This is related in a deep
way to the accessory parameters of the associated fuchsian
equation \cite{takhtajan}.

For completeness let us now turn to the effective action for the
breathing mode of the internal manifold (see also
\cite{porrati,Aghababaie,randjbar}). Depending on the values of
the parameters this mode might be as heavy as the first KK modes
in which case it should be integrated out. It is however important
to check that the mass is positive so that the compactification is
stable. This is not guaranteed in general. To derive the effective
action we consider the following ansatz for the metric,
\begin{equation}
ds^2=\phi^{-2}(x)g_{\mu\nu}(x)dx^\mu dx^\nu+\phi^2(x)
\psi(z,\bar{z})dz d\bar{z}
\end{equation}
Conservation of the flux requires that $F$ remains at its ground
state value (\ref{flux}). Plugging the ansatz into the action and
using the Liouville equation for the background we obtain,
\begin{equation}
S_4=M_6^4\int \frac{\psi} 2 dz d\bar{z} \int d^4 x \sqrt{-g}
\left(\frac {R_4} 2-2 \frac{\partial^\mu \phi
\partial_\mu \phi} {\phi^2}-V\right)
\label{s4}
\end{equation}
where,
\begin{equation}
V=\frac {\lambda} {\phi^2}-\left(\frac {\lambda} 2+\frac 3 4
B_0^2\right)\frac 1 {\phi^4}+\frac {B_0^2} {2 \phi^6}
\end{equation}
By means of the volume formula (\ref{volume}) the four dimensional
Planck mass is,
\begin{equation}
M_4^2=M_6^4 V_2=M_6^4 \frac {2 \pi} {k}(2-2g-\sum_i \alpha_i)
\end{equation}
Notice that from the low energy point of view the only effect of
the branes is to change the normalization of the Planck mass. It
should be stressed that, as can be seen from (\ref{s4}), the KK
reduction is consistent so that no tadpoles corrections arise to
the classical effective action.

As required the potential has a stationary point at $\phi=1$ which
corresponds to dS, AdS or Minkowski space according to
(\ref{cases}). The mass of $\phi$ is given by,
\begin{equation}
m_{\phi}^2=\frac 3 2 B_0^2- \lambda
\end{equation}
We conclude that the compactification is stable unless
$\lambda>3/2 B_0^2$ which corresponds to dS space (see also
\cite{santiago}). In this case the system will roll to the other
stationary point of the potential at $\phi^2=3B_0^2/(2\lambda)$.

\section{Conclusions}
\label{conclusions}

Let us summarize what we have achieved in this paper. Starting
from the football shaped extra dimensions scenario with two equal
tension branes \cite{carroll,navarro}, we have generalized the
model to include an arbitrary number of branes. We have also
considered the case where the ground state is dS or AdS space and
the internal manifold is a Riemann surface. The internal space has
constant curvature with conical singularities at the location of
the branes. The problem of determining the metric consists in
finding a solution of the Liouville equation with singularities, a
topic which goes back to Poincar\'e and Picard. Explicit solutions
have been presented for the case of three branes. Most
importantly, contrary to the scenario with two branes, the
tensions of the branes do not need to be tuned with each other but
only satisfy mild constraints. For the case relevant to the
Minkowski background, topologically the internal space is a
sphere. For three branes (say $T_3\ge T_2\ge T_1$) solutions exist
when,
\begin{eqnarray}
T_1+T_2+T_3&<& 4\pi M_6^4\nonumber \\
T_3&<&T_1+T_2
\end{eqnarray}
where the first condition is a direct consequence of the
Gauss-Bonnet theorem while the second has a more mysterious
geometrical origin. We conjectured in (\ref{nbranes}) the
generalization of this formula to the scenario with an arbitrary
number of branes. Finally we have described the low energy
effective action for the model. For more than three branes, the
positions of the branes are not fixed in the ground state so $N-3$
complex moduli appear in the low energy effective theory.

\section*{Acknowledgments}
I am grateful to Massimo Porrati for very helpful discussions
about the Liouville equation and Riemann surfaces. I would also
like to thank Massimo Porrati and especially Jose Santiago for
comments on the manuscript. This work was supported in part by the
NSF grant PHY-0245068.

\appendix
\section*{Riemann-Hilbert problem}

The solution of the Liouville equation is closely related to the
Riemann-Hilbert problem of determining functions with prescribed
monodromies in the complex plane.

In order to solve this problem one introduces the fuchsian
equation,
\begin{equation}
\frac {d^2 u} {dz^2}+\sum_{i=1}^{N-1}\left[\frac
{\alpha_i(2-\alpha_i)}{4(z-z_i)^2}+\frac
{c_i}{2(z-z_i)}\right]u=0, \label{fuchapp}
\end{equation}
where $\alpha_i$ are directly related to the monodromies and $c_i$
are known as the accessory parameters. The condition that infinity
is a regular singular point of the fuchsian equation implies three
linear equations on the $c_i$'s,
\begin{equation}
\left\{
\begin{array}{l}
\sum_{i=1}^{N-1} c_i = 0 \\
\sum_{i=1}^{N-1} \left[ 2 c_i z_i + \alpha_i(2-\alpha_i) \right] = \alpha_{\infty}(2-\alpha_{\infty}) \\
\sum_{i=1}^{N-1} \left[ c_i z_i^2 + z_i (\alpha_i(2-\alpha_i))
\right] = c_{\infty}
\end{array}
\right. \label{accessory}
\end{equation}
so that the $c_i$ are fully determined for $N=3$. The double poles
singularities in (\ref{fuchapp}) fix the behavior of the solutions
near the singular points,
\begin{equation}
u(z)\sim A\, (z-z_i)^{1-\frac {\alpha_i} 2}+B\, (z-z_i)^{\frac
{\alpha_i} 2}.
\end{equation}
from which one can easily derive the monodromies. Given a pair of
linearly independent solutions ($u_1$, $u_2$), it is easy to see
that $w=u_1/u_2$ satisfies the Liouville equation
(\ref{liouville}). However, since the monodromy of ($u_1$, $u_2$)
belongs in general to $SL(2,\mathbb{C})$, the function $\psi$ is
not single valued. In order to find a solution well defined on the
entire complex plane one needs to require that the monodromies are
contained in $SU(2)$ (or $SU(1,1)$ when the curvature is
negative). These conditions determine the accessory parameters
$c_n$ as well as ($u_1$, $u_2$). Since $w=u_1/u_2$ now transforms
as,
\begin{equation}
w\to \frac {a w+ b} {-\bar{b} w+
\bar{a}},~~~~~~~~~~~~~|a|^2+|b|^2=1,
\end{equation}
it leaves (\ref{solution}) invariant. Therefore $w$ defines a
single valued solution of the Liouville equation on the complex
plane.

As a simplest example one can consider the case with two
singularities. Using reparametrization invariance these can be
chosen at $(0, \infty)$. The constraints (\ref{accessory})
determine the fuchsian equation to be,
\begin{equation}
\frac {d^2 u} {dz^2}+\frac {\alpha_1(2-\alpha_1)}{4 z^2}\,u=0.
\end{equation}
Notice that eqs. (\ref{accessory}) also requires
$\alpha_1=\alpha_\infty$. Two linearly independent solutions are,
\begin{eqnarray}
u_1&=&z^{1-\frac {\alpha_1} 2}\nonumber \\
u_2&=&z^{\frac {\alpha_1} 2}.
\end{eqnarray}
Since the monodromy of these solutions is obviously contained in
$SU(2)$, $w=u_1/u_2$ is a well defined solution of the Liouville
equation. In fact this just reproduces the football solution
(\ref{2branes}). This derivation also shows that there are no
other solutions with two branes.

\label{appendix}

\end{document}